\documentclass[nofootinbib]{revtex4}
\usepackage{amsmath}
\usepackage{amsfonts}
\usepackage{amssymb}
\usepackage{latexsym}
\usepackage{graphicx}
\usepackage[english]{babel}
\input epsf.sty

\begin{document}

\title{
{Null Energy Condition and Dark Energy Models}}
 \vspace{3mm}
\author{Taotao Qiu\footnote{qiutt@mail.ihep.ac.cn}, Yi-Fu
Cai\footnote{caiyf@mail.ihep.ac.cn} and Xinmin Zhang}

\affiliation{Institute of High Energy Physics, Chinese Academy of
Sciences, P.O. Box 918-4, Beijing 100049, P. R. China}

\begin{abstract}
Null Energy Condition (NEC) requires the equation of state (EoS) of
the universe $w_u$ satisfy $w_u\geq-1$, which implies, for instance
in a universe with matter and dark energy dominating
$w_u=w_m\Omega_m+w_{de}\Omega_{de}=w_{de}\Omega_{de}\geq-1$. In this
paper we study constraints on the dark energy models from the
requirement of the NEC. We will show that with $\Omega_{de}\sim0.7$,
$w_{de}<-1$ at present epoch is possible. However, NEC excludes the
possibility of $w_{de}<-1$ forever as happened in the Phantom model,
but if $w_{de}<-1$ stays for a short period of time as predicted in
the Quintom theory NEC can be satisfied. We take three examples of
Quintom models of dark energy, namely the phenomenological EoS, the
two-scalar-field model and the single scalar model with a modified
Dirac-Born-Infeld (DBI) lagrangian to show how this happens.
\end{abstract}

\maketitle

\section{Introduction}

It is well known that energy conditions play an important role in
classical theory of general relativity and
thermodynamics\cite{Hawking:1973uf}. In classical general relativity
it is usually convenient and efficient to restrict a physical system
to satisfy one or some of energy conditions for study, for example,
in the proof of Hawking-Penrose singularity
theorem\cite{Penrose:1964wq,Hawking:1969sw}, the positive mass
theorem\cite{Schon:1981vd} and so on, while in thermodynamics energy
conditions are the bases for obtaining entropy
bounds\cite{Bousso:1999xy,Flanagan:1999jp}. Among those energy
conditions, the null energy condition is the weakest one which
states that for any null vector $n^\mu$ the stress energy tensor
$T_{\mu\nu}$ should satisfy the relation
\begin{eqnarray}
T_{\mu\nu} n^\mu n^\nu \geq 0~.
\end{eqnarray}

In general, the violation of NEC leads to the breakdown of causality
in general relativity and the violation of the second law of
thermodynamics\cite{ArkaniHamed:2007ky}. These pathologies require
that the total stress tensor in a physical spacetime manifold should
obey the NEC. In the framework of the standard 4-dimensional
Friedmann-Robertson-Walker (FRW) cosmology the NEC implies
$\rho+p\geq 0$, which in turn gives rise to a constraint on the
equation of state of the universe (EoS) $w_u$ defined as the ratio
of pressure to energy density, $w_{u}\geq -1$. In this paper we
study the constraints on the dark energy models from the requirement
of $w_{u}\geq -1$.

In the early Universe with radiation dominant
the EoS of the universe
$w_{u}$ is approximately equal to $\frac{1}{3}$ and in the matter
dominant period $w_{u}$
is nearly zero, so NEC is satisfied well. However when the dark energy
component is not negligible we have
\begin{eqnarray}\label{NECd}
w_u=w_m\Omega_m+w_{de}\Omega_{de}\geq-1~,
\end{eqnarray}
where the subscripts `$m$' and `$de$' stand for matter and dark
energy, respectively. With $w_m = 0$, inequality (\ref{NECd})
becomes
\begin{eqnarray}\label{NECde}
w_{de}\Omega_{de}\geq-1~.
\end{eqnarray}
From the inequality above, we can see that models of dark energy
with $w_{de}\geq-1$ such as the Cosmological Constant and the
Quintessence satisfy the NEC, but the models with $w_{de}<-1$
predicted for instance by the Phantom theory where the kinetic
term of the scalar field has a wrong sign does not. Interestingly
we can see that NEC might be satisfied in models if $w_{de}<-1$
stays for a short period of time during the evolution of the
universe. In this paper we will show this happens in the Quintom
models of dark energy.

The paper is organized as follows: in section II we will present
three examples of the Quintom models to show how the NEC is
satisfied and the section III is the summary of the paper.

\section{Null Energy Condition and the Quintom Dark
Energy}

Quintom is a dynamical model of dark energy\cite{Quintom1}. It
differs from the Cosmological Constant,
Quintessence\cite{Ratra:1987rm}, Phantom\cite{Caldwell:1999ew},
K-essence\cite{Chiba:1999ka} and so on in the determination of the
cosmological evolution. Although the current data in combination
with the 3-year WMAP\cite{Spergel:2006hy}, the recently released 182
SNIa Gold sample\cite{Riess06} and also other cosmological
observational data show the consistence of the Cosmological
Constant, it is worth noting that the dynamical dark energy models
are not excluded and Quintom dark energy is mildly favored (for
recent references see e.g. \cite{Zhao:2006qg,Lihong2006,Zhao2006}).
The most salient feature of the Quintom model is that its EoS can
smoothly cross $-1$. In this section we will study the implications
of NEC on the Quintom models. Working with three specific examples
we will show the NEC can indeed be satisfied.

\subsection{ A phenomenological model with parameterized EoS across $-1$ }

With a simple calculation, the inequality (\ref{NECde}) can be
rewritten as
\begin{eqnarray}\label{NECdew}
\left(1+w_{de}(a)\right)\Omega_{de0}\exp\left\{\int_1^a[-3(1+w_{de}(a'))]d\ln
a'\right\}+\frac{\Omega_{m0}}{a^3}\geq0~,
\end{eqnarray}
where the subscript `0' represents today's value.

We firstly start with a phenomenological model with a parameterized
EoS which will be able to cross over $-1$:
\begin{equation}\label{para1}
w_{de}=w_0+w_1 (1 - a )~,
\end{equation}
where $a$ is the scale factor which we normalize to be $a_0 =1$ at
present. One can see that when $a$ is equal to $(1+w_0+w_1)/w_1$,
the EoS of dark energy crosses $-1$. This type of parametrization
for the EoS has been widely used in the
literature\cite{Chevallier:2000qy,Linder:2002et}for the fitting of
constraining the EoS to the observational data.

\begin{figure}[htbp]
\includegraphics[scale=1.0]{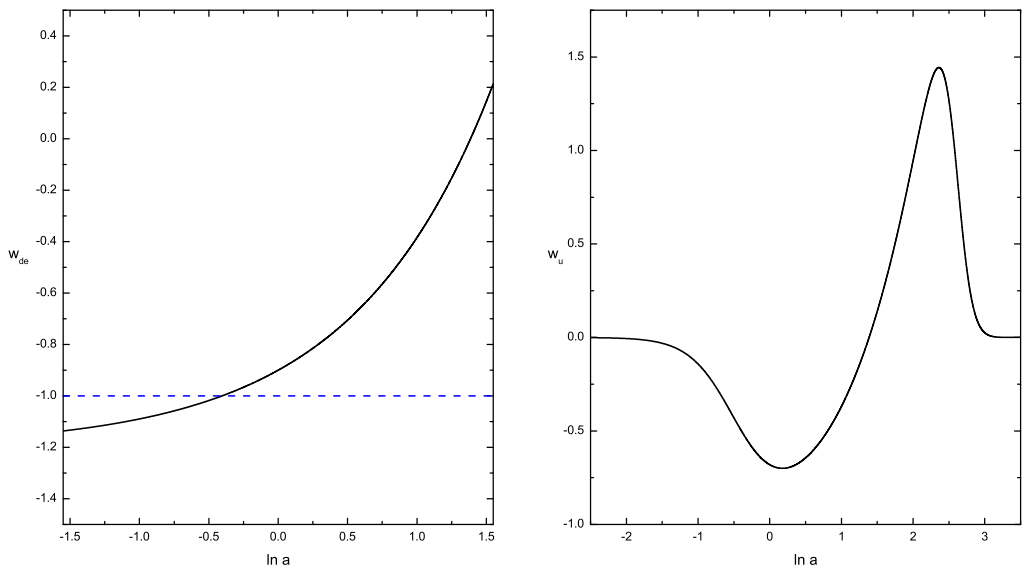}
\caption{Plot of the evolution of the EoS $w_{de}$
with $w_{de}=w_0+w_1(1-a)$,
and the EoS $w_u$ for the universe as a function of $\ln a$. Here
in the numerical calculation we have taken $w_0=-0.9$ and
$w_1=-0.3$.} \label{fig1:para}
\end{figure}

Inserting Eq. (\ref{para1}) into the inequality (\ref{NECdew}), the NEC
puts a constraint on the parameters
$w_0$
and $w_1$. In Fig. \ref{fig1:para}, we take
$w_0=-0.9$ and $w_1=-0.3$, and then plot the
evolution of the dark energy EoS  and also the EoS of the universe
respectively. One can see from Fig. \ref{fig1:para} that the EoS
of dark energy $w_{de}$ crosses $-1$ from below to above.
In this case the EoS of the universe $w_u$ evolves from zero which
corresponds to the matter dominant epoch, and then reaches its
minimal value during which the universe enters the dark energy
dominant period, and finally returns to zero in the future. We can
read directly from this figure that the minimal value of $w_u$
stays above $-1$ which satisfies the NEC.

\begin{figure}[htbp]
\includegraphics[scale=1.0]{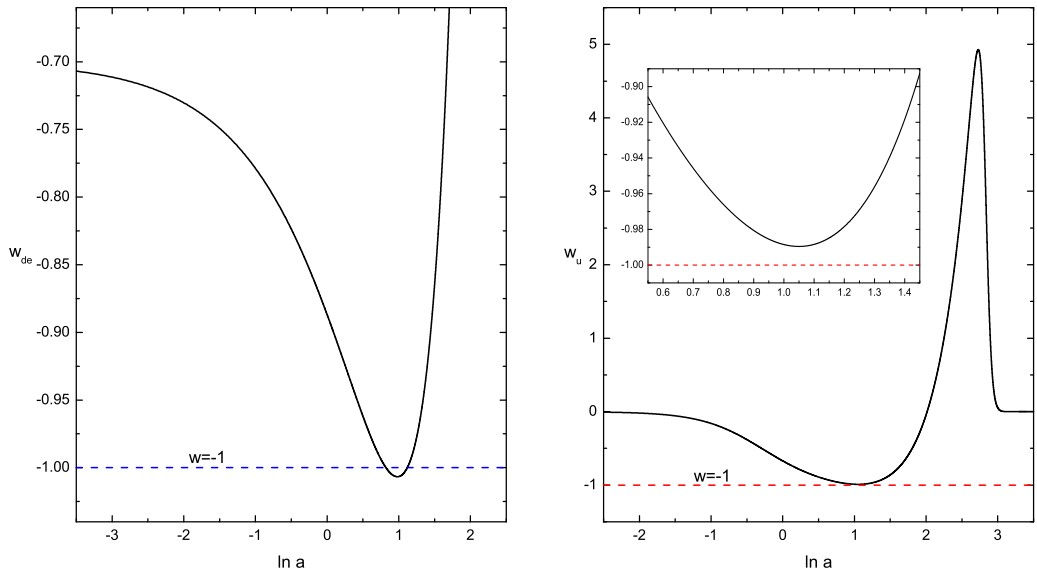}
\caption{Plot of the evolution of the EoS $w_{de}$ 
with
$w_{de}=w_0+w_1(1-a+\frac{3}{16}a^2)$, and the EoS $w_u$ for the
universe as a function of $\ln a$. Here in the numerical calculation
we have taken $w_0=-0.93$ and $w_1=0.23$.} \label{fig2:para}
\end{figure}

One could take another example with a different parametrization of the
EoS:
\begin{equation}\label{para2}
w_{de}=w_0+w_1(1-a+\frac{3}{16}a^2)~.
\end{equation}
This example is different from the previous one since we introduce a
square term of the scale factor which makes the EoS $w_{de}$ crosses
$-1$ twice. Taking proper values of those parameters, this model can
give a scenario that dark energy stays in the Phantom-like state
only for a while and then returns to be Quintessence-like. After
taking the similar calculation, we plot the evolution of the EoS of
the model and the universe respectively in Fig. \ref{fig2:para}. In
this example the EoS of dark energy $w_{de}$ starts to evolve from
Quintessence-like in the past, and then enters the Phantom-like
state for a short period of time, and finally returns to above $-1$.
We also find that $w_u$ behaves similar to that in Fig.
\ref{fig1:para}. Its value resides on zero in the past and in the
future which means that matter have dominated and will dominate
again the evolution of the universe. Although $w_u$ runs away from
zero during the evolution, its value keeps being larger than $-1$ as
shown in Fig. \ref{fig2:para}. Therefore, this model can be
consistent with the NEC as well.

\subsection{The Two-scalar-field Quintom model}

Building the field model of Quintom dark energy is a challenge due
to the No-Go theorem which has been proved in Ref.
\cite{xiacai07}(also see Ref.
\cite{Quintom1,zhaogbper,Caldwell05,Vikman05,wHu05,Kunz06}). This
No-Go theorem forbids a traditional scalar field model with a
lagrangian of general form ${\cal L}={\cal L}(\phi,
\nabla_\mu\phi\nabla^\mu\phi)$ from having its EoS cross over the
cosmological constant boundary. According to this theorem,
dynamical models like Quintessence, Phantom and K-essence are
unable to realize their EoS cross $-1$. Therefore, to realize a
viable Quintom field model in the framework of Einstein's gravity
theory, one needs to introduce extra degrees of freedom to the
conventional theory with a single scalar field. The simplest
Quintom model is constructed by two scalars with one being
Quintessence-like and another Phantom-like proposed firstly in
Ref. \cite{Quintom1}, and this model has been widely studied in
detail later on. In recent years there have been a lot of
activities in the theoretical study on building Quintom models,
such as a single scalar with high-derivative \cite{lfz,arefeva},
vector field\cite{Wei}, extended theory of
gravity\cite{extendgravity}, Lorentz-violating dark energy
models\cite{Huang:2005gu} and so on, see e.g. \cite{Quintomsum}.

In this section we investigate a two-scalar-field Quintom model of
dark energy in flat FRW cosmology which is described by the action
\begin{eqnarray}
S=\int d^4x \sqrt{-g} \left[\frac{R}{16\pi G}+{\cal L}_m+{\cal
L}_{de}\right]~,
\end{eqnarray}
with
\begin{eqnarray}\label{lagrangede}
{\cal L}_{de}=\frac{1}{2}\partial_{\mu}\phi_1\partial^{\mu}
\phi_1-\frac{1}{2}\partial_{\mu}\phi_2\partial^{\mu}\phi_2-V(\phi_1,\phi_2)~,
\end{eqnarray}
where $R$ is the Ricci scalar of the universe, ${\cal L}_m$ is the
lagrangian of matter, ${\cal L}_{de}$ is the lagrangian of dark
energy, and the metric is in form of $(+,-,-,-)$. Here the field
$\phi_1$ has a canonical kinetic term, but $\phi_2$ is a ghost
field. With ${\cal L}_{de}$ in (\ref{lagrangede}), we can easily
obtain the energy density and the pressure of this model,
\begin{eqnarray}
\rho_{de}=\frac{1}{2}{\dot\phi_1}^2-\frac{1}{2}{\dot\phi_2}^2+V~,\\
p_{de}=\frac{1}{2}{\dot\phi_1}^2-\frac{1}{2}{\dot\phi_2}^2-V~,
\end{eqnarray}
where the dot denotes the derivative with respect to the cosmic
time, and by the variational principle the Einstein equations are
given by
\begin{eqnarray}
\label{Einstein1}
H^2=\frac{8\pi G}{3}(\frac{1}{2}{\dot\phi_1}^2-\frac{1}{2}{\dot\phi_2}^2+V+\rho_m)~,\\
\label{Einstein2}
\ddot\phi_1+3H\dot\phi_1+\frac{dV}{d\phi_1}=0~,\\
\label{Einstein3} \ddot\phi_2+3H\dot\phi_2-\frac{dV}{d\phi_2}=0~.
\end{eqnarray}
Since it is required that the total EoS of the universe satisfies
the NEC, it is clear that the Phantom component of this kind of
model can only be dominant for a short while and hence it gives a
constraint on choosing the potential of the Phantom field. In the
following numerical calculations, we will choose an appropriate
potential of Phantom field and different potentials of Quintessence
field, which can satisfy the NEC and give various Quintom scenarios.

\begin{figure}[htbp]
\includegraphics[scale=1.0]{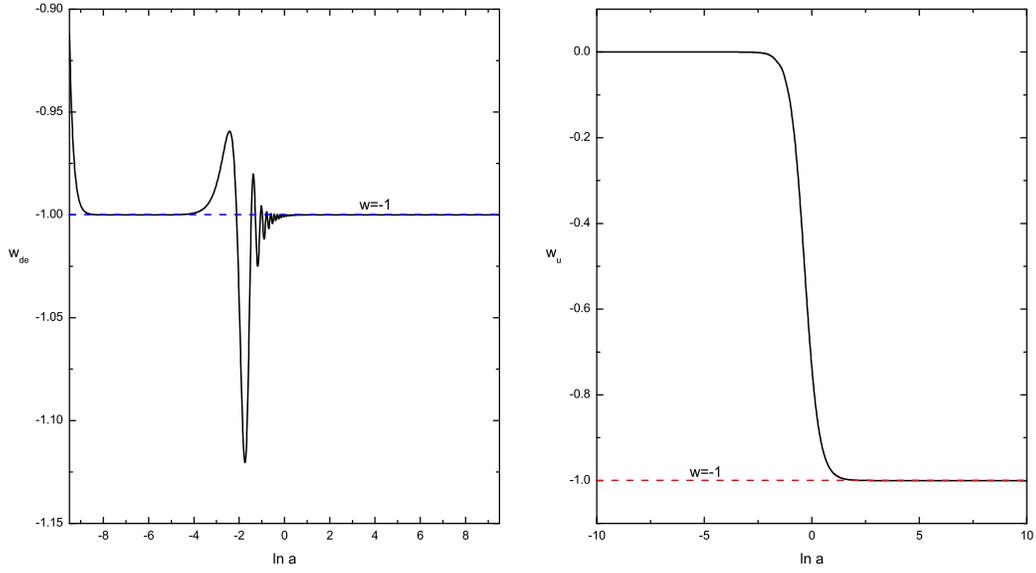}
\caption{Plot of the evolution of the EoS $w_{de}$ for the
two-scalar-field Quintom dark energy with the potential
$V_{\phi_1}=V_1\exp(\bar\lambda_1\phi_1^2),~
V_{\phi_2}=V_2\exp(\bar\lambda_2\phi_2^2)$, and the EoS $w_u$ for
the universe as a function of $\ln a$. Here in the numerical
calculation we have taken $V_1=(3.1623eV)^4$, $V_2=(1.8612eV)^4$,
$\bar\lambda_1=-0.8\times 10^4$, $\bar\lambda_2=-0.5\times 10^4$
and the initial values are $\phi_{1i}=0.02$,
$\dot\phi_{1i}=5.0\times 10^{-62}$, $\phi_{2i}=-0.01$ and
$\dot\phi_{2i}=2.0\times 10^{-62}$.} \label{fig3:double}
\end{figure}

In Fig. \ref{fig3:double}, we choose the potentials of
Quintessence and Phantom field to be
$V_{\phi_1}=V_1\exp(\lambda_1\phi_1^2/M^2)$ and
$V_{\phi_2}=V_2\exp(\lambda_2\phi_2^2/M^2)$, respectively. Here we
normalize the fields with Planck scale and choose $M$ to be
$0.01M_{pl}$.\footnote{Here and in the following, we would like to
redefine the parameter $\bar\lambda_1$ and $\bar\lambda_2$ to be
$\lambda_1/M^2$ and $\lambda_2/M^2$.} During the evolution, the
Phantom field will climb up along its potential and become
dominant because its energy density will increase while that of
Quintessence field will decrease. However, since there is a
maximum of the potential, after reaching that point, the Phantom
field will stay on the top and stop affecting the evolvement of
the universe. From the figure, we can see that after the EoS of
dark energy being less than $-1$ today for a short while, it can
exit in the future and actually approach to $-1$. In this case we
read from the figure that the EoS of the universe can be always
larger than $-1$, and thus the NEC can be easily satisfied.

\begin{figure}[htbp]
\includegraphics[scale=1.0]{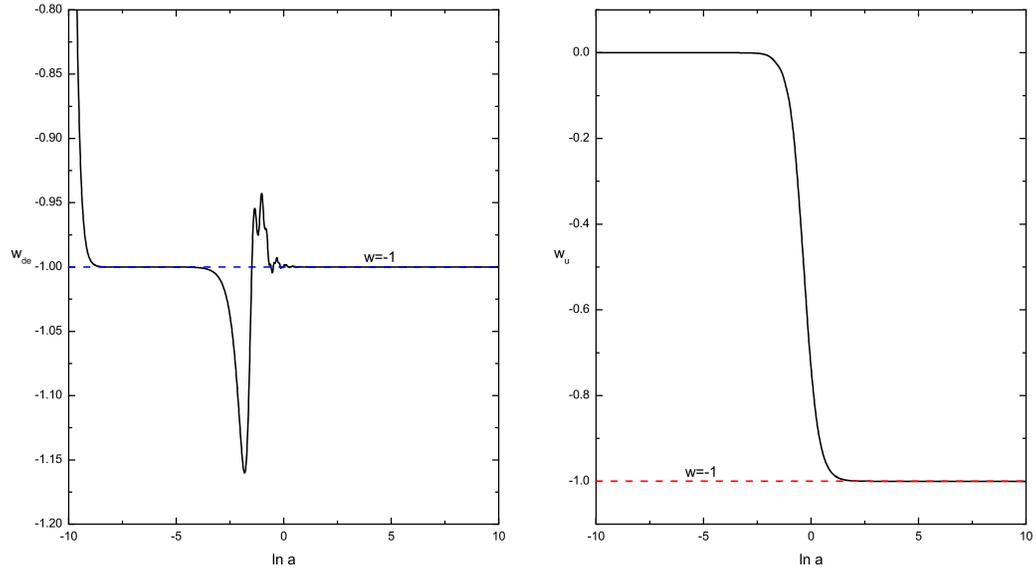}
\caption{Plot of the evolution of the EoS $w_{de}$ for the
two-scalar-field Quintom dark energy with the potential
$V_{\phi_1}=\frac{1}{2}m^2\phi_1^2,~
V_{\phi_2}=V_2\exp(\bar\lambda_2\phi_2^2)$,  and the EoS $w_u$ for
the universe as a function of $\ln a$. Here in the numerical
calculation we have taken $m=1.0\times 10^{-29}eV$,
$V_2=(1.8612eV)^4$, $\bar\lambda_2=-0.5\times 10^4$ and the
initial values are $\phi_{1i}=0.02$, $\dot\phi_{1i}=5.0\times
10^{-62}$, $\phi_{2i}=-0.01$ and $\dot\phi_{2i}=2.0\times
10^{-62}$.} \label{fig4:double}
\end{figure}

In Fig. \ref{fig4:double}, we choose another potential form of the
Quintessence field to be $V_{\phi_1}=\frac{1}{2}m^2\phi_1^2$ with
the Phantom's potential unchanged. We can see from the figure
that, similar to the reason above, Phantom can only dominate the
universe for a while, and due to different potential of
Quintessence, the EoS of dark energy model behaves very
differently. The figure shows that after the universe exit the
Phantom dominating phase, the Quintessence will dominate the
universe again, and the EoS of dark energy will cross $-1$ twice.
In the future, however, this EoS will also become de-sitter like.
In the whole process, the EoS of the universe will always be above
the cosmological constant boundary, and this model of Quintom dark
energy also satisfies the NEC.

\subsection{A single scalar field with a modified DBI lagrangian}

Having presented the examples of two-scalar-field Quintom models in
consistent with the NEC, we in this section consider a class of
Quintom models described by an effective lagrangian involving higher
derivative operators.

Due to the contribution of higher derivative terms, this kind of
models can give rise to an EoS across $-1$ as pointed out in Ref.
\cite{lfz}. A connection of this type of Quintom theory to the
string theory has been considered in Ref. \cite{cyftachyon} and
\cite{arefeva}. Here we take the string-inspired model in
\cite{cyftachyon} for a detailed study to check whether it satisfies
the requirement of the NEC. The action of this Quintom dark energy
is given by
\begin{eqnarray}\label{tachyonaction}
S_{de}=\int d^4x\sqrt{-g}\left[-V(\phi)\sqrt{1-{\alpha}^\prime
\nabla_{\mu}\phi\nabla^{\mu}\phi+{\beta}^\prime\phi\Box\phi}\right]~.
\end{eqnarray}
This is a generalized version of ``Born-Infeld"
action\cite{Gerasimov:2000zp, Kutasov:2000qp} with the introduction
of the $\beta^\prime$ term(see, for example, \cite{Cheung:2004sa}
for derivation of the higher derivative term from string theory). To
the lowest order, the Box-operator term $\phi\Box\phi$ is equivalent
to the term $\nabla_{\mu}\phi\nabla^{\mu}\phi$ when the tachyon is
on the top of its potential. However when the tachyon rolls down
from the top of the potential, these two terms exhibit different
dynamical behavior. The two parameters $\alpha'$ and $\beta'$ in
(\ref{tachyonaction}) could be arbitrary in the case of the
background flux being turned on \cite{Mukhopadhyay:2002en}. One
interesting feature of this model is that it provides the
possibility of its EoS $w_{de}$ running across the cosmological
constant boundary. In the following studies to make two parameters
($\alpha^\prime$, $\beta^\prime$) dimensionless, it is convenient to
redefine $ \alpha =\alpha^\prime M^4$ and $\beta = \beta^\prime M^4$
where $M$ is an energy scale of the effective theory of the field.

From (\ref{tachyonaction}) we obtain the equation of motion for
the scalar field $\phi$:
\begin{eqnarray}\label{eqom}
\frac{\beta}{2}\Box(\frac{V\phi}{f})+
\alpha\nabla_{\mu}(\frac{V\nabla^{\mu}\phi}{f})+M^4V_{\phi}f+\frac{\beta
V}{2f}\Box\phi=0~,
\end{eqnarray}
where $f=\sqrt{1-{\alpha}^\prime
\nabla_{\mu}\phi\nabla^{\mu}\phi+{\beta}^\prime\phi\Box\phi}$ and
$V_\phi=dV /d\phi$. Correspondingly, the energy stress tensor of
Quintom dark energy is given by
\begin{eqnarray}\label{stress}
T_{\mu\nu}&=&g_{\mu\nu}[Vf-\frac{\beta}{2M^4}\nabla_{\rho}(\frac{\phi
V}{f}\nabla^{\rho}\phi)]+
\frac{\alpha}{M^4}\frac{V}{f}\nabla_\mu\phi\nabla_\nu\phi+\frac{\beta}{2M^4}\nabla_\mu(\frac{\phi
V}{f})\nabla_\nu\phi\nonumber\\
&&+\frac{\beta}{2M^4}\nabla_\nu(\frac{\phi V}{f})\nabla_\mu\phi~.
\end{eqnarray}
In order to simplify the calculation, we technically define
another parameter $\psi\equiv\frac{\partial{\cal
L}}{\partial\Box\phi}=-\frac{\beta\phi V}{2M^4f}$ to solve
(\ref{eqom}) and (\ref{stress}). In the framework of a flat FRW
universe filled with a homogeneous scalar field $\phi$, we have
the equations of motion in forms of
\begin{eqnarray}
\label{tachyon1}\ddot\phi+3H\dot\phi&=&\frac{\beta\phi}{4M^4\psi^2}V^2-\frac{M^4}{\beta\phi}+\frac{\alpha}{\beta\phi}\dot\phi^2~,\\
\label{tachyon2}\ddot\psi+3H\dot\psi&=&(2\alpha+\beta)(\frac{M^4\psi}{\beta^2\phi^2}-\frac{V^2}{4M^4\psi})-\frac{\beta\phi}{2M^4\psi}VV_{\phi}
-\frac{2\alpha}{\beta\phi}\dot\psi\dot\phi\nonumber\\
&&-(2\alpha-\beta)\frac{\alpha\psi}{\beta^2\phi^2}\dot\phi^2~,
\end{eqnarray}
and the energy density and the pressure of this field can be written
as
\begin{eqnarray}
\label{rho}\rho_{de}=-\frac{\alpha\psi}{\beta\phi}\dot\phi^2-\dot\psi\dot\phi-\frac{\beta\phi}{4M^4\psi}V^2-\frac{M^4\psi}{\beta\phi}~,\\
\label{press}
p_{de}=-\frac{\alpha\psi}{\beta\phi}\dot\phi^2-\dot\psi\dot\phi+\frac{\beta\phi}{4M^4\psi}V^2+\frac{M^4\psi}{\beta\phi}~.
\end{eqnarray}

According to the restriction of the NEC, we need this Quintom model
to satisfy the inequality (\ref{NECdew}). Although in this case the
phase space of Quintom dark energy is constrained, we will show
below that the inequality (\ref{NECdew}) can be satisfied easily. In
the numerical study, we constrain the parameters $\alpha$ and
$\beta$ so that when expanding the derivative terms in the square
root to the lowest order the model in (\ref{tachyonaction}) gives
rise to a canonical kinetic term for the scalar field $\phi$
\cite{cyftachyon}, i.e., $\alpha+\beta>0$.

\begin{figure}[htbp]
\includegraphics[scale=1.0]{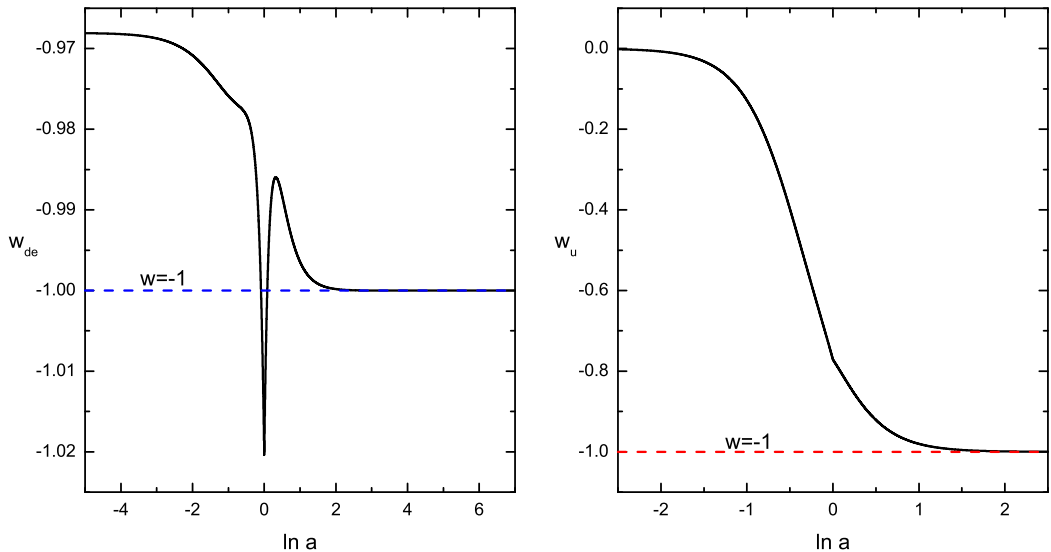}
\caption{Plot of the evolution of the EoS $w_{de}$ for the
string-inspired Quintom dark energy with the potential
$V(\phi)=\frac{V_0}{\exp(-\lambda\phi)+\exp(\lambda\phi)}$ and the
EoS $w_u$ for the universe as a function of $\ln a$. Here in the
numerical calculation we have taken $\lambda=10^{-3}$,
$V_0=(1.3183\times 10^{-3}eV)^4$, $\alpha=0.6$, $\beta=0.4$ and
the initial values are $\phi_i=12$, $\phi_i'=-2.3\times 10^{-5}$,
$\psi_i=-3.84\times 10^2$ and $\psi_i'=-40$. In choosing the
initial values the prime represents the derivative with respect to
$\ln a$.} \label{fig5:dbi}
\end{figure}

Through calculating Eqs. (\ref{tachyon1}), (\ref{tachyon2}) and
the Friedmann equations numerically and then comparing the results
in the inequality (\ref{NECdew}), we can judge whether a model is
consistent with the NEC. We first consider a model with the
potential
$V(\phi)=\frac{V_0}{\exp(-\lambda\phi)+\exp(\lambda\phi)}$. In the
calculation we normalize the fields $\phi$ and $\psi$ with the
Planck scale and choose the energy scale $M=1.2211\times
10^{-4}eV$. For a given set of the model parameters, we make the
numerical calculations and plot the evolutions of the EoS of the
Quintom dark energy and the universe, which are shown in Fig.
\ref{fig5:dbi}. We can read that the EoS of dark energy $w_{de}$
starts from a value larger than $-1$ then evolves to less than
$-1$, and soon exits the Phantom-like state, and eventually
approaches the cosmological constant asymptotically. During the
evolution the period of time for dark energy to be Phantom-like is
very short.
 Therefore, the corresponding EoS of the universe evolves
from the matter dominant period with $w_u=0$ to the dark energy
dominant period smoothly without violating the NEC as shown in Fig.
\ref{fig5:dbi}.

\begin{figure}[htbp]
\includegraphics[scale=1.0]{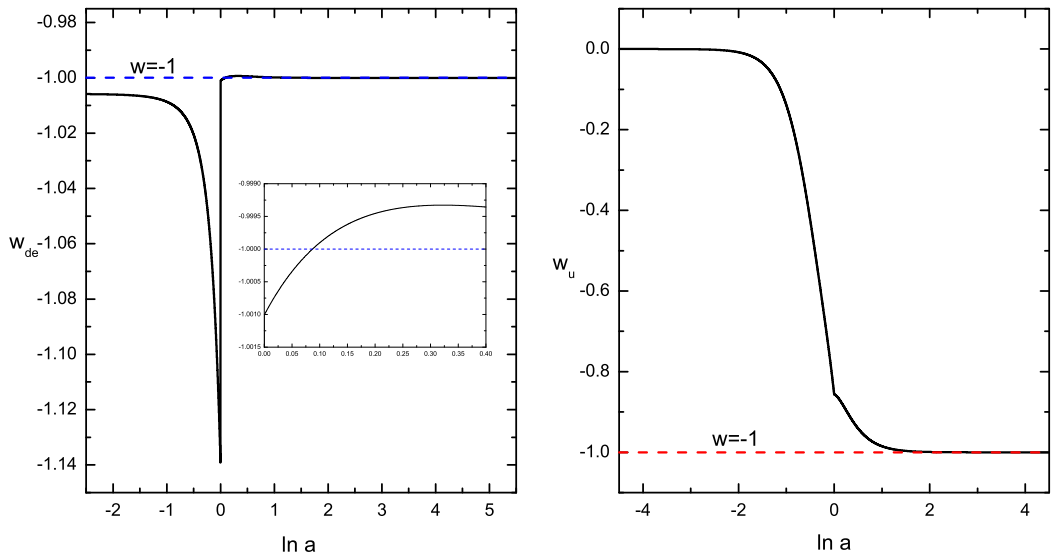}
\caption{Plot of the evolution of the EoS $w_{de}$ for the
string-inspired Quintom dark energy with the potential
$V(\phi)=\frac{1}{2}m^2\phi^2$ and the EoS $w_u$ for the universe
as a function of $\ln a$. Here in the numerical calculation we
have taken $m=1.435\times 10^{-35} eV$, $\alpha=1.3$, $\beta=0.5$
and the initial values are $\phi_i=10$, $\phi_i'=-4.3\times
10^{-3}$, $\psi_i=-4\times 10^2$ and $\psi_i'=-1.9$. In choosing
the initial values the prime represents the derivative with
respect to $\ln a$.} \label{fig6:dbi}
\end{figure}

For another example we take a different form of potential
$V(\phi)=\frac{1}{2}m^2\phi^2$. Similar to what have been done
above, we obtain another example consistent with the NEC. The
normalization in this case is the same as that in the above one. In
Fig. \ref{fig6:dbi}, we give the evolutions of the EoS of this
Quintom model and the universe respectively. One can see from this
figure the $w_u$ satisfies the NEC, but the detailed evolution of
the universe differs from the one shown in Fig. \ref{fig5:dbi}. In
Fig. \ref{fig6:dbi} the EoS of Quintom dark energy $w_{de}$ crosses
the cosmological constant boundary from a fixed value below $-1$ to
above and then evolves close to $-1$ in the future. In this case the
EoS of the universe $w_u$ runs from zero when the universe is
dominated by the matter, and then approaches the de-Sitter phase
asymptotically.

\section{Conclusion}

In this paper we have studied the implications of NEC in the models
of dark energy. We show that NEC excludes the models with
$w_{de}<-1$ forever as predicted by the Phantom dark energy, however
allows the possibility of having $w_{de}<-1$ for a short period of
time as it happens in the Quintom models. We have shown explicitly
in this paper three examples of Quintom models where NEC is
satisfied.

\section*{Acknowledgments}

We thank Tom Banks, Rong-Gen Cai, Yun-Song Piao and Jun-Qing Xia for
useful discussions. This work is supported in part by National
Natural Science Foundation of China under Grant Nos. 90303004,
10533010 and 10675136 and by the Chinese Academy of Science under
Grant No. KJCX3-SYW-N2.


\end{document}